\documentclass{PoS}
\usepackage{graphicx}
\usepackage{amsmath}
\usepackage{txfonts}
\usepackage{amssymb}
\usepackage{subfig}
\usepackage{cite}

\title{Host galaxies of double-peaked $\left[ \textrm{OIII} \right]$ emitting AGN: binary AGN or mergers?}

\ShortTitle{Host galaxies of double-peaked $\left[ \textrm{OIII} \right]$ emitting AGN}

\author{\speaker{Carolin Villforth}\thanks{}\\
        Department of Astronomy, University of Florida, Gainesville, FL 32611-2055, USA\\
        E-mail: \email{villforth@astro.ufl.edu}}
\author{Fred Hamann\\
       	Department of Astronomy, University of Florida, Gainesville, FL 32611-2055, USA\\}
        
\abstract{Mergers are suspected to be reliable triggers of both starformation and AGN activity. However, the exact timing of this process remains poorly understood. Here, we present deep imaging and long slit spectroscopy data of a sample of four double-peaked $\left[ \textrm{OIII} \right]$ emitting AGN. These sources are often believed to host binary AGN, or at least be currently undergoing major mergers. The sample presented here either have previous IFU and high resolution imaging data that show double-nuclei in the IR as well as kinematicly and spatially distinct line emitting regions.  Two sources have detections of double point sources in either the X-ray or radio. The sources studied are therefore likely binary AGN. The AGN in this sample are luminous, radio-quiet and at low redshift. The $u/r/z$ imaging data show host galaxies in a wide range of merger stages, with the majority (3/4) showing tidal tails or complex kinematics and morphologies clearly indicating a recent merger. One source however -- hosting a double X-ray source-- shows quiescent morphologies with no clear signs of interaction from imaging data. The spectroscopy in this case reveals a gas disk counter-rotating with respect to the stellar component. Spectroscopy of the other sources reveal disturbed kinematics further confirming their status as ongoing mergers. Our data show that AGN triggering in mergers may happen over a wide time span and that sinking of black holes to the center of a merged system might take considerable time in some cases. A detailed analysis will be published in an upcoming paper. Further studies of AGN in merging galaxies will show how the hosts of those AGN differ from normal mergers without AGN activity.}

\FullConference{The Extreme sky: Sampling the Universe above 10 keV - extremesky2009,\\
		November 6-8, 2012\\
		Max-Planck-Insitut f\"ur Radioastronomie (MPIfR), Bonn,  			Germany}

\begin{document}

%\section{Introduction}

%The connection between massive gas-rich mergers and Active Galactic Nuclei (AGN) has been the source of much debate for several decades \cite{sanders_ultraluminous_1988,hopkins_cosmological_2008}. It is clear that gas-rich mergers provide an ideal environment for trigering both massive starburst and AGN activity \cite{hopkins_cosmological_2008}. Observational evidence for a strong connection between mergers and AGN activity has remained mixed \cite{bennert_evidence_2008,cisternas_bulk_2011,cisternas_secular_2011,kocevski_candels:_2012,villforth_seds_2012}.

%\cite{abazajian_seventh_2009}

\section{Sample \& Data}

The sample was chosen from double-peaked $\left[ \textrm{OIII} \right]$ emitters from the literature \cite{smith_search_2010}, observable from the GTC in the fall (see Table \ref{T:objects}). Three of the sources (SDSS0952,SDSS1151,SDSS1502) have double-peaked $\left[ \textrm{OIII} \right]$ emission lines and double stellar cores in the IR \cite{fu_mergers_2011,fu_nature_2012,shen_type_2011}. SDSS1502 additionally shows two compact radio cores inconsistent with starformation \cite{fu_kiloparsec-scale_2011}. The fourth binary AGN in our sample (SDSS171544.0+600835) is also a double-peaked $\left[ \textrm{OIII} \right]$ emitter, but was followed in the X-rays with Chandra \cite{comerford_chandra_2011}, it shows double nuclei, best explained as a sign of binary AGN. Data were taken at the 10.4m Gran Telescopio Canarias (GTC) on La Palma (Canary Islands, Spain) during dark time. Observing conditions were excellent with clear sky conditions and subarcsecond seeing during all observations (0.56-0.75" in the z band).  All observations were obtained with OSIRIS, an imager and spectrograph for the optical wavelength range, located in the Nasmyth-B focus of GTC. We obtained $u$/$r$/$z$ imaging as well as long-slit optical medium resolution (R=2500,$\lambda$=5630-7540\AA) spectroscopy for all objects. All data were reduced using standard IRAF routines. Photometric calibration was performed using field stars.

\section{Results \& Conclusions}

\begin{figure*}
\includegraphics[width=8cm]{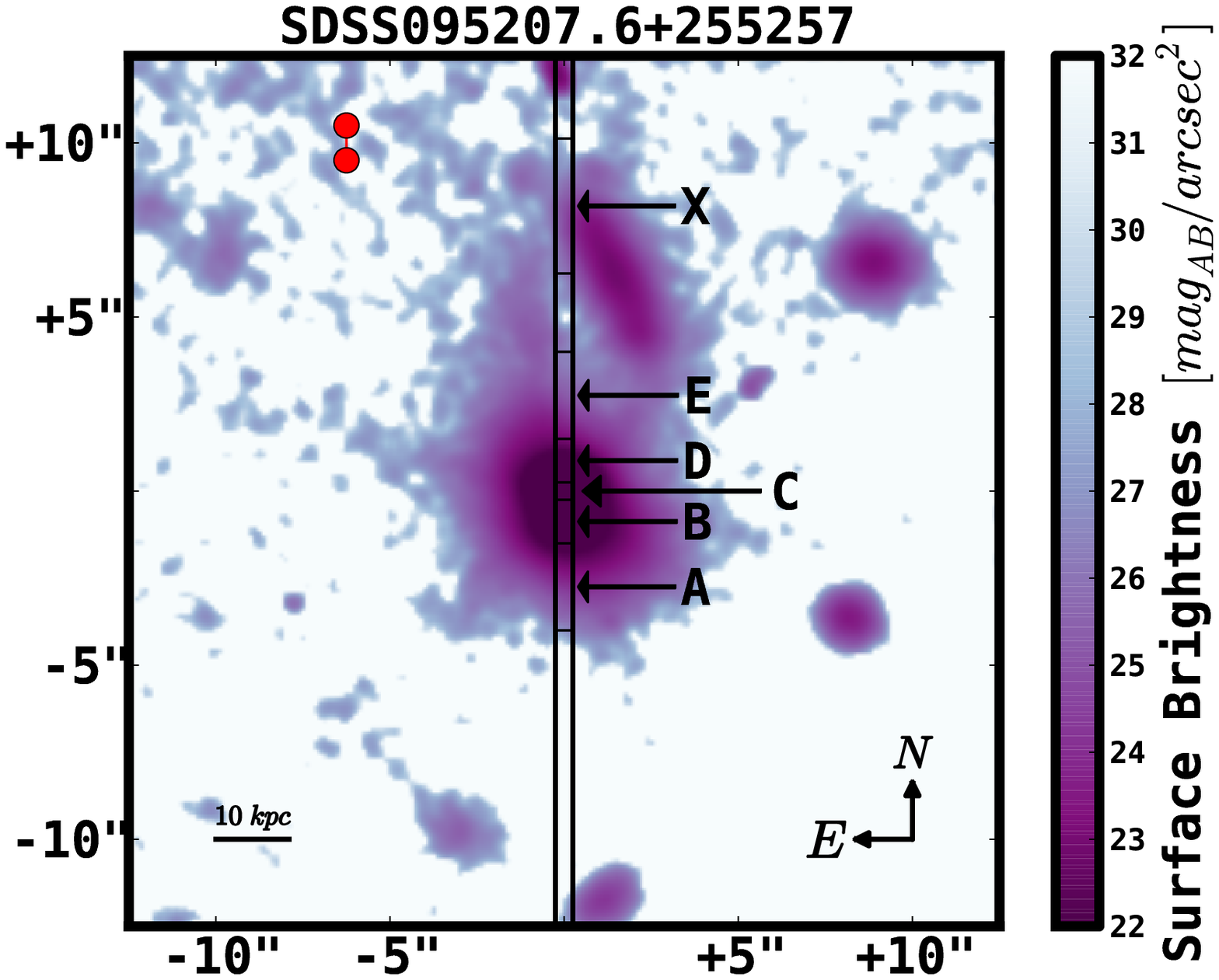}
\includegraphics[width=8cm]{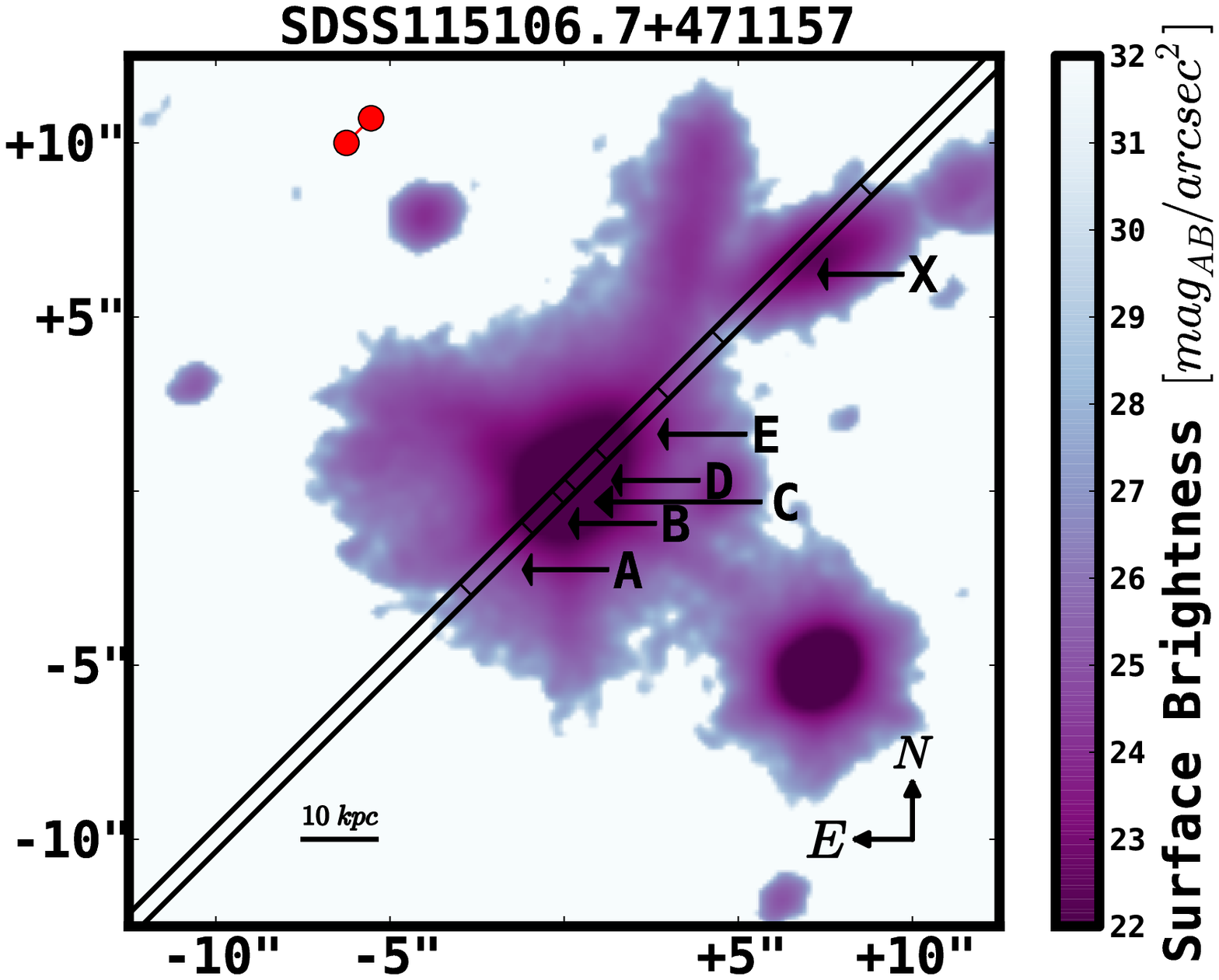}
\includegraphics[width=8cm]{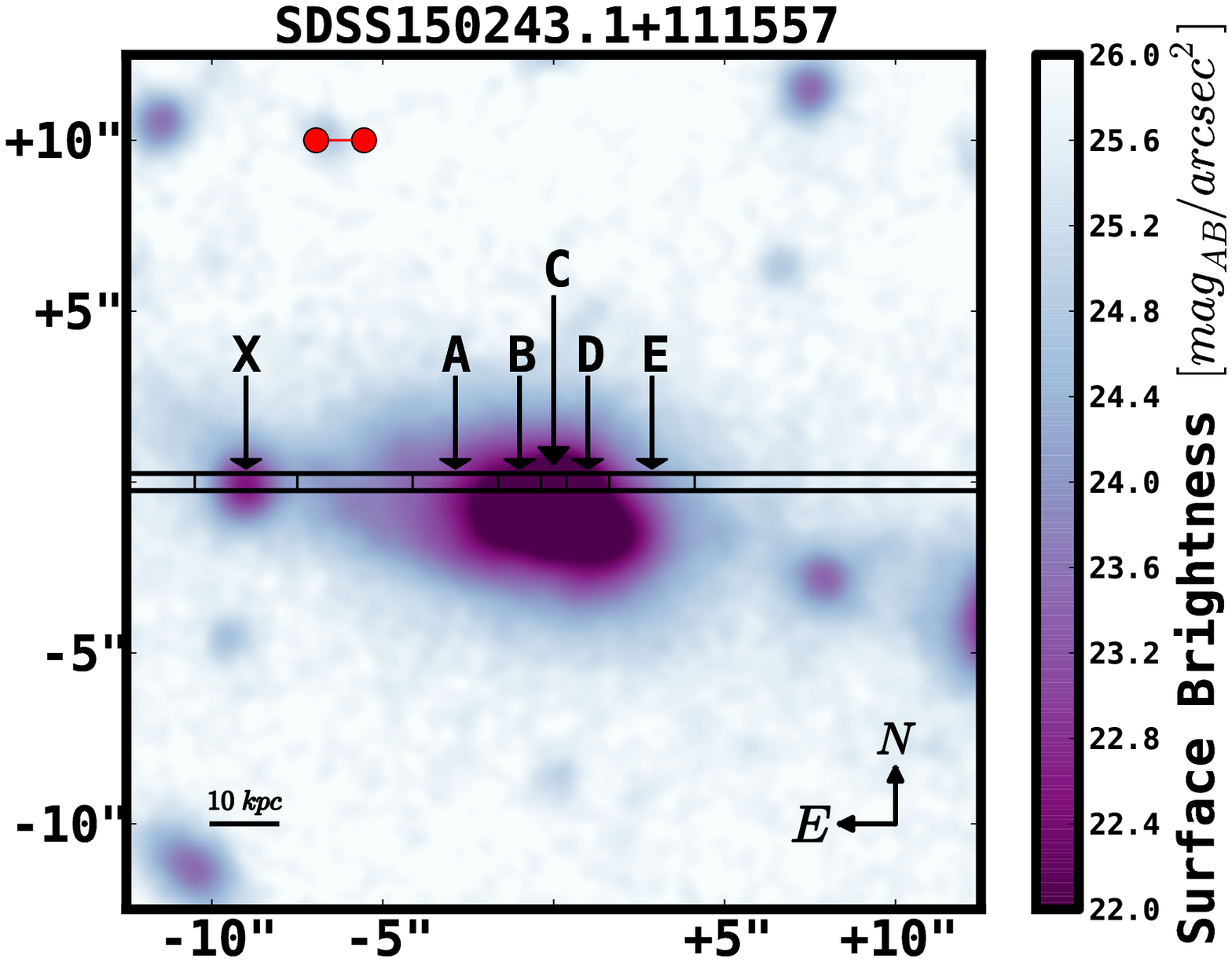}
\includegraphics[width=8cm]{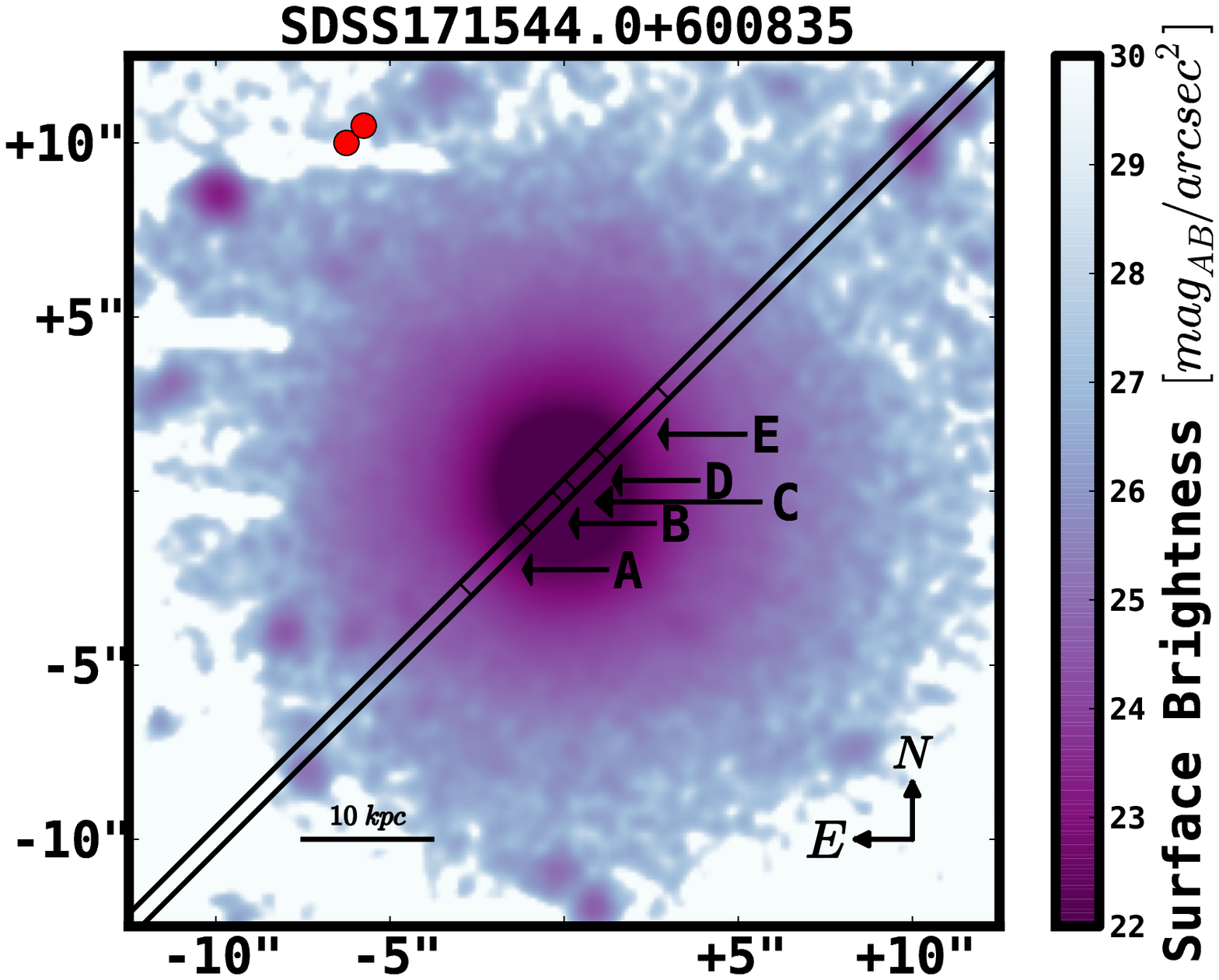}
\caption{$r$ band surface brightness maps of sources studied. All plots are 25$\times$25; north is up and east is left.  A 10kpc scale is shown in all four plots. The separations and orientations of the two cores are shown as red filled circles in the upper left corner. The black lines and labels show the long-slit position used for spectroscopy and slices along which spectra were extracted. The dominant Type 1 AGN is always located at position C.}
\label{F:slitmaps}
\end{figure*}

The $r$ band surface brightness maps for all sources in our sample are shown in Fig. \ref{F:slitmaps}. We see a wide range of merger properties, though three of the four sources studied clearly are in a very early stage of the merger, with tidal tails and complex morphologies.

\begin{figure*}
\begin{center}
\includegraphics[width=7.5cm]{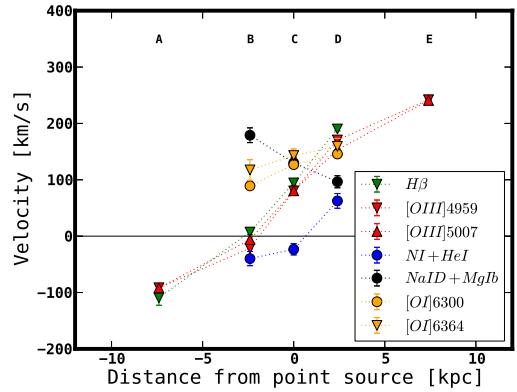}
\caption{Kinematic structure of SDSS1715. Left: velocities are from line fits to extracted spectra in slices shown in Fig. 1. The velocities are calculated with respect to the catalogued SDSS catalogue redshifts.}
\label{F:kinematics}
\end{center}
\end{figure*}

An exception is SDSS1715 -- identified as a binary AGN in the X-rays \cite{comerford_chandra_2011} -- which shows a quiescent host galaxy with no clear signs of disturbance down to low surface brightness limits. The host galaxy is best fit as a disk/bulge mixture. The kinematics of the source however is very interesting (see Fig. \ref{F:kinematics}), the gas in the objects shows rotation along the slit, the stellar component (detected in NaID as well as MgIb) shows rotation as well, albeit with a different direction. This is similar to cases of galaxies where stellar and gas disks are misaligned \cite{morse_1998,coccato_2011,coccato_2013}, though such cases do not commonly appear in larger surveys of gas kinematics in galaxies (e.g. \cite{dumas_2007}). Such a kinematic structure can be interpreted as a sign of a recent merger of a gas-dominated component with angular momentum misaligned with the stellar disk. It is interesting that this is observed in SDSS1715, which host a double X-ray source at a separation of only $\sim$ 1.9kpc \cite{comerford_chandra_2011}. Since the event in which the counter-rotating gas disk was acquired did not disturbed the stellar disk, it is likely also not the source of the second black hole. This implies that the binary black hole was formed in an earlier event. Sinking of the black holes to the center must therefore have taken a considerable time since no singns of disturbance are visible.

Our results show that AGN in mergers can be observed in a wide range of merger stages and that in some cases, such as SDSS1715, the sinking of black holes to the center of a merged galaxy can take till long after merger features have disappeared. Detailed analysis of the full dataset will be presented in an upcoming paper.

\begin{table}
\caption{Binary AGN sample properties and observation information. ID: SDSS name, z: redshift, $\theta ["]$: separation between AGN in arcseconds, $\theta [kpc]$ between AGN in kpc; log(R): log of radio-loudness.}
\begin{tabular}{ccccccc}
\hline
ID & Type & z  & $log(L_{[OIII] 5007} [erg/s])$ & $\theta$ [``] & $\theta$ [kpc] & log(R) \\
\hline
SDSS095207.6+255257 & Type1/2 & 0.339 & 42.34 & 0.99 & 4.9 &  <0.27\\
SDSS115106.7+471157 & Type1/2 & 0.318 & 43.20 & 1 & 4.7  &  0.75\\
SDSS150243.1+111557	& Type1/2 & 0.390 & 43.35 & 1.39 & 7.4  & 1.8\\
SDSS171544.0+600835 & Type2/2 (X-ray) & 0.157 & 42.23  & 0.68  & 1.85 & 1.95\\
\hline
\end{tabular}
\label{T:objects}
\end{table}

\begin{thebibliography}{99}
\bibitem{bennert_evidence_2008} Bennert et al. 2008, ApJ, 677, 846
\bibitem{cisternas_bulk_2011} Cisternas et al. 2011, ApJ, 726, 57
\bibitem{cisternas_secular_2011} Cisternas et al. 2011, ApJ Letters, 741, 11
\bibitem{coccato_2011} 	Coccato et al. 2011, MNRAS Letters, 412, 113
\bibitem{coccato_2013} 	Coccato et al. 2013, A\& A, 549, 3
\bibitem{comerford_chandra_2011} Comerford et al. 2011, ApJ Letters, 737, 19
\bibitem{dumas_2007} Dumas et al. 2007, MNRAS, 379, 1249
\bibitem{fu_mergers_2011} Fu et al. 2011, ApJ, 733, 103
\bibitem{fu_nature_2012} Fu et al. 	2012, ApJ, 745, 67
\bibitem{fu_kiloparsec-scale_2011} Fu et al. 2011, ApJ Letters 740, 44
\bibitem{hopkins_cosmological_2008} Hopkins et al. 2008, ApJS, 175, 356
\bibitem{kocevski_candels:_2012} 	Kocevski et al. 2012, ApJ, 744,148
\bibitem{morse_1998} Morse et al.  1998, ApJ, 505, 159
\bibitem{sanders_ultraluminous_1988} Sanders et al. 1988, ApJ, 325, 74
\bibitem{shen_type_2011} Shen at al. 2011, ApJ, 735, 48
\bibitem{smith_search_2010} Smith et al. 2010, ApJ, 716, 866
\bibitem{villforth_seds_2012} Villforth et al. 2012, MNRAS, 426, 360
\end{thebibliography}
\end{document}